\documentclass[aps,prc,twocolumn,superscriptaddress]{revtex4-1}
\usepackage{CJK}
\usepackage{graphicx}% Include figure files
\usepackage{hyperref}
\usepackage{amssymb}
\usepackage{amsmath}
\usepackage[normalem]{ulem}
\usepackage{url}
\usepackage{color}

\begin{document}
\begin{CJK*} {UTF8}{} %{GB} {gbsn}
%\preprint{APS/123-QED}

\title{Extracting Skyrme energy density functional parameters with Heavy-Ion Collision Data}

\author{Yingxun Zhang} %(\CJKfamily{gbsn} ÕÅӢѷ)} %张英é€)
\affiliation{China Institute of Atomic Energy, Beijing 102413, P.R. China}
\email{zhyx@ciae.ac.cn}
\affiliation{Guangxi Key Laboratory Breeding Base of Nuclear Physics and Technology, Guilin 541004, China}

%\author{M. Pierre} %(\CJKfamily{bsmi} 曾敏å…}
%\affiliation{National Superconducting Cyclotron Laboratory and Physics and Astronomy Department, Michigan State University, East Lansing, MI 48824, USA}

%\author{G. Jhang}
%\affiliation{National Superconducting Cyclotron Laboratory and Physics and Astronomy Department, Michigan State University, East Lansing, MI 48824, USA}

%\author{Y.C. Tsang}
%\affiliation{National Superconducting Cyclotron Laboratory and Physics and Astronomy Department, Michigan State University, East Lansing, MI 48824, USA}

%\author{Hang Liu} % (\CJKfamily{gbsn} 刘航 }
%\affiliation{Texas Advanced Computing Center, University of Texas, Austin, Texas 78758, USA}

\author{Zhuxia Li} %(\CJKfamily{gbsn} 李祝éœ}
\affiliation{China Institute of Atomic Energy, Beijing 102413, P.R. China}
\author{Min Liu} %(\CJKfamily{gbsn} ÕÅӢѷ)} %张英é€)
\affiliation{Guangxi Normal University, Guilin, 541004, P.R.China}
%\author{Ning Wang} %(\CJKfamily{gbsn} ÕÅӢѷ)} %张英é€)
%\affiliation{Guangxi Normal University, Guilin, 541004, P.R.China}
%\author{M.B.Tsang} %(\CJKfamily{bsmi} 曾敏å…}
%\affiliation{National Superconducting Cyclotron Laboratory and Physics and Astronomy Department, Michigan State University, East Lansing, MI 48824, USA}

\date{\today}

\begin{abstract}
The effective Skyrme energy density functionals are widely used in the study of nuclear structure, nuclear reaction and neutron star, but they are less established from the heavy ion collision data. In this work, we find 22 effective Skyrme parameter sets, when incorporated in use the transport model, ImQMD, describe the heavy ion collision data, such as isospin diffusion data at 35 MeV/u and 50 MeV/u. We use these sets to calculate the neutron skin of $^{208}$Pb based on the restricted density variation method, and obtain the neutron skin of $^{208}$Pb in the range of $\delta R_{np}=0.18\pm0.04$ fm.
\end{abstract}

\pacs{21.60.Jz, 21.65.Ef, 24.10.Lx, 25.70.-z}
%\keywords{}

\maketitle
\end{CJK*}

\emph{Introduction}. Over last couple decades, the effective nuclear energy density functionals, which take into account some of the complicated correlations that characterize complex nuclei, have become a tool for describing the properties of nuclear ground-states, heavy ion collisions and neutron stars.
Considerable progress has been achieved in constructing and optimizing the effective energy density functionals (EDF), both in nonrelativistic\cite{Skyrme56,Vauthe72,Chab97,Decharge80,Klu09,Korte10,Blaiz95,Goriely09} and relativistic\cite{Brock78,Horo83, Bouy87,WHLong06,Niksik08,Lala05,PWZhao10} frameworks. Recently, construction of the EDFs has been also inspired by ab initio calculations \cite{Dobac16} and effective field theories\cite{Bonnard18,ZZhang18}.

Traditionally, the effective Skyrme energy density functionals are more commonly used in nuclear structure, reactions and astrophysics studies for their relative simplicity in computation but containing sufficient physics to allow quantitative description of structures and reactions of nuclei\cite{Greiner96}. The parameters in the Skyrme energy density functionals are obtained through best fitting of the nuclear matter parameters, as well as the properties of nucleus, such as binding energy, shell gap, rms radii, fission barriers and giant resonance energies, some of them also consider the properties of neutron stars. So far, more than 200 parameter sets with their corresponding nuclear matter parameters have been obtained. As shown in reference\cite{Dutra12}, the uncertainties in predictions of nuclear matter parameters from the compiled parameter sets, such as, incompressibility $K_0=9\rho^2 \frac{\partial^2 \epsilon/\rho}{\partial \rho^2}|_{\rho_0}$, isoscalar effective mass $\frac{m}{m_s^*}=(1+\frac{2m}{\hbar^2}\frac{\partial}{\partial \tau}\frac{E}{A})|_{\rho_0}$\cite{Klu09}, symmetry energy coefficient $S_0=S(\rho_0)$, slope of symmetry energy $L=3\rho_0\frac{\partial S(\rho)}{\partial \rho}|_{\rho_0}$, and isovector effective mass $\frac{m_v^*}{m}=\frac{1}{1+\kappa}$, where $\kappa$ is the enhancement factor of the Thomas-Reich-Kuhn sum rule\cite{PRing80}, still exits.
%In the following discussions, we replace $m^*_v$ by $f_I$ which is defined as $f_I=\frac{1}{2\delta}(\frac{m}{m_n^*}-\frac{m}{m_p^*})=\frac{m}{m_s^*}-\frac{m}{m_v^*}$. And the correlation between them is not clear and depends on the data they used to describe.
One method of improving the Skyrme energy density functional is to constrain it in a multi-dimensional parameter space and in a large density range, which can be realized by best fitting the heavy ion collision data with the transport model calculations. Heavy ion collision can form high density in its compression phase, and subnormal density during its expansion, thus, the heavy ion collision can check the effective energy density functional over a large density region. In another, one has to remove a $prior$ correlations on the nuclear matter parameters when one use the data to obtain the effective Skyrme energy density functional parameters.

In this work, we adopt the nuclear matter parameters \{$K_0$, $S_0$, $L$, $m_s^*$, $f_I$ \} as independent inputs and then to obtain the effective Skyrme interaction parameter sets. Here, we replace $m^*_v$ by $f_I$, which is defined as $f_I=\frac{1}{2\delta}(\frac{m}{m_n^*}-\frac{m}{m_p^*})=\frac{m}{m_s^*}-\frac{m}{m_v^*}$, since the $f_I$ can be analytically incorporated into the transport model and its sign reflects the $m^*_n>m_p^*$ or $m^*_n<m_p^*$.
The range of nuclear matter parameters \{$K_0$, $S_0$, $L$, $m_s^*$, $f_I$ \} and the correlation between them are estimated, and 22 effective Skyrme parameter sets, are obtained by comparing the transport model calculations to the HIC data, such as isospin diffusion data at 35 and 50 MeV/u, are discussed. Finally, we use the obtained 22 parameter sets to calculate the neutron skin of $^{208}$Pb using the restricted density variational method.

\emph{ImQMD model}. The transport model used in this work is the ImQMD-Sky\cite{Zhang14,Zhang15}. In the model, the nucleonic potential energy density without the spin-orbit term is $u_{loc}+u_{md}$, and
\begin{eqnarray}
\label{eq:edfimqmd}
u_{loc}=&&\frac{\alpha}{2}\frac{\rho^2}{\rho_0} +\frac{\beta}{\eta+1}\frac{\rho^{\eta+1}}{\rho_0^\eta}+\frac{g_{sur}}{2\rho_0 }(\nabla \rho)^2+\\\nonumber
&&\frac{g_{sur,iso}}{\rho_0}[\nabla(\rho_n-\rho_p)]^2+A_{sym}\frac{\rho^2}{\rho_0}\delta^2+B_{sym}\frac{\rho^{\eta+1}}{\rho_0^\eta}\delta^2
\end{eqnarray}
and Skyrme-type momentum dependent energy density functional $u_{md}$ is written based on its interaction form $\delta (r_1-r_2 ) (p_1-p_2 )^2$\cite{Skyrme56, Vauthe72,Zhang15} as,
\begin{eqnarray}
\label{eq:mdimqmd}
u_{md}=&&C_0\sum_{ij}\int d^3pd^3p' f_i(r,p)f_j(r,p')(p-p')^2+\\\nonumber
&&D_0\sum_{ij\in n}\int d^3pd^3p'f_i(r,p) f_j(r,p')(p-p')^2 +\\\nonumber
&&D_0\sum_{ij\in p}\int d^3p d^3p' f_i(r,p)f_j(r,p')(p-p')^2.
\end{eqnarray}
The connection between 9 parameters $\alpha$, $\beta$, $\eta$, $A_{sym}$, $B_{sym}$, $C_0$, $D_0$, $g_{sur}$, $g_{sur,iso}$ used in ImQMD-Sky and the 9 nuclear matter parameters, \{$\rho_0, E_0,K_0,S_0,L,m_s^*,m_v^*, g_{sur}, g_{sur,iso}$\}, are given by following analytical relationship,
\begin{eqnarray}
&& g_{\rho\tau}=\frac{3}{5}(\frac{m_0}{m_s^*}-1)\epsilon_F^0, \\\nonumber &&\eta=(K_0+\frac{6}{5}\epsilon_F^0-10g_{\rho\tau})/(\frac{9}{5}\epsilon_F^0-6g_{\rho\tau}-9E_0)\\\nonumber
&&\beta=\frac{(\frac{1}{5}\epsilon_F^0-\frac{2}{3} g_{\rho\tau}-E_0 )(\eta+1)}{\eta-1}, \alpha=E_0-\epsilon_F^0-\frac{8}{3} g_{\rho\tau}-\beta,\\\nonumber
&& C_0=\frac{1}{16\hbar^2}\Theta_v, D_0=\frac{1}{16\hbar^2}(\Theta_s-2\Theta_v),\\\nonumber
&&C_{sym}=-\frac{1}{24}(\frac{3\pi^2}{2})^{2/3} (3\Theta_v-2\Theta_s )\rho_0^{5/3},\\\nonumber
&&B_{sym}=\frac{3S_0-L-\frac{1}{3}\epsilon_F^0+2C_{sym} (m_s^*,m_v^* )}{-3\sigma}\\\nonumber
&&A_{sym}=S_0-\frac{1}{3}\epsilon_F^0-B_{sym}-C_{sym} (m_s^*,m_v^*)
\label{eq:skyqmd}
\end{eqnarray}
where $\Theta_s=(\frac{m_0}{m_s^*}-1) \frac{8\hbar^2}{m_0\rho_0}$, $\Theta_v=(\frac{m_0}{m_v^*}-1)\frac{4\hbar^2}{m_0\rho_0}$, and $\eta=\sigma+1$. Similar relation has been discussed in references\cite{Agrawal05,LWChen09}. The novel approach used in this work is that we set the 9 nuclear matter parameters \{$\rho_0, E_0,K_0,S_0,L,m_s^*,m_v^*, g_{sur}, g_{sur,iso}$ \} as the input of the ImQMD-Sky code. The coefficients of surface terms are set as $g_{sur} = 24.5MeV fm^2$ and $g_{sur;iso} = -4.99MeV fm^2$, and varying of $g_{sur}$ and $g_{sur,iso}$ in a reasonable region for different Skyrme interactions has negligible effects on the calculated experimental observables in intermediate energy heavy ion collisions. The nucleon-nucleon collision and Pauli-blocking part used in this work are treated as the same as that in Ref\cite{Zhang05,Zhang06,Zhang07}, and we do not vary its strength or form in this study since previous calculations have shown it does not strongly influence the isospin sensitive observables we studied\cite{Zhang12}.

\emph{Density variational method}. The approach we used to calculate the neutron skin is the restricted density variational method (RDV), which as the same as in the Ref.\cite{MLiu06}, where the semi-classical expressions of the Skyrme energy density functional are applied to study the ground state of energies, the neutron proton density distributions, and the neutron skin thickness of a series of nuclei. The binding energy of a nucleus is expressed as the intergral of energy density fucntional, i.e.
\begin{equation}
\label{eq:edf}
E=\int \mathcal{H} dr=\int \frac{\hbar^2}{2m}[\tau_n(\mathbf{r})+\tau_p(\mathbf{r})]+\mathcal{H}_{sky}+\mathcal{H}_{coul} dr
\end{equation}
The energy density functional $\mathcal{H}_{sky}$ is nucleonic density functional, which has the same form as we used in the ImQMD model, but with the spin-orbit interaction form.
%One has to notice, the kinetic energy density are given by,
%\begin{eqnarray}
%\label{eq:tau}
%\tau_i(\mathbf{r})&=&\frac{3}{5}(3\pi^2)^{2/3}\rho_i^{5/3}+\frac{1}{36}\frac{(\nabla\rho_i)^2}{\rho_i}+\frac{1}{3}\triangle\rho_i\\\nonumber
%&&+\frac{1}{6}\frac{\nabla\rho_i+\nabla f_i+rho_i\triangle f_i}{f_i}-\frac{1}{12}\rho_i(\frac{\nabla f_i}{f_i})^2\\\nonumber
%&&+\frac{1}{2}\rho_i(\frac{2m}{\hbar^2}\frac{W_0}{2}\frac{\nabla(\rho+\rho_i)}{f_i})^2,
%\end{eqnarray}
%where the extended Thomas-Fermi (ETF) approach including all terms up to second order (ETF2) and forth order (ETF4) as in Ref.\cite{Brack85}. $\rho_i$ denotes the proton and neutron density of nucleus, and $\rho=\rho_n+\rho_o$. $W_0$ is the strength of the spin-orbit interaction; the parameter $f_i(\mathbf{f})$ is as the same as in reference\cite{MLiu06}. The Coulomb energy density is written as the sum of the direct and exchange terms.
In our calculations, we take the density distribution as a spherical symmetric Fermi function:
\begin{equation}
\rho_i=\rho_{0i}[1+exp(\frac{r-R_{0i}}{a})], i={n, p}.
\end{equation}
By minimizing the total energy of the system given by Eq.~(\ref{eq:edf}), the neutron and proton densities can be obtained and thus the neutron skin.

\begin{figure}[htbp]
\centering
\includegraphics[angle=0,scale=0.28]{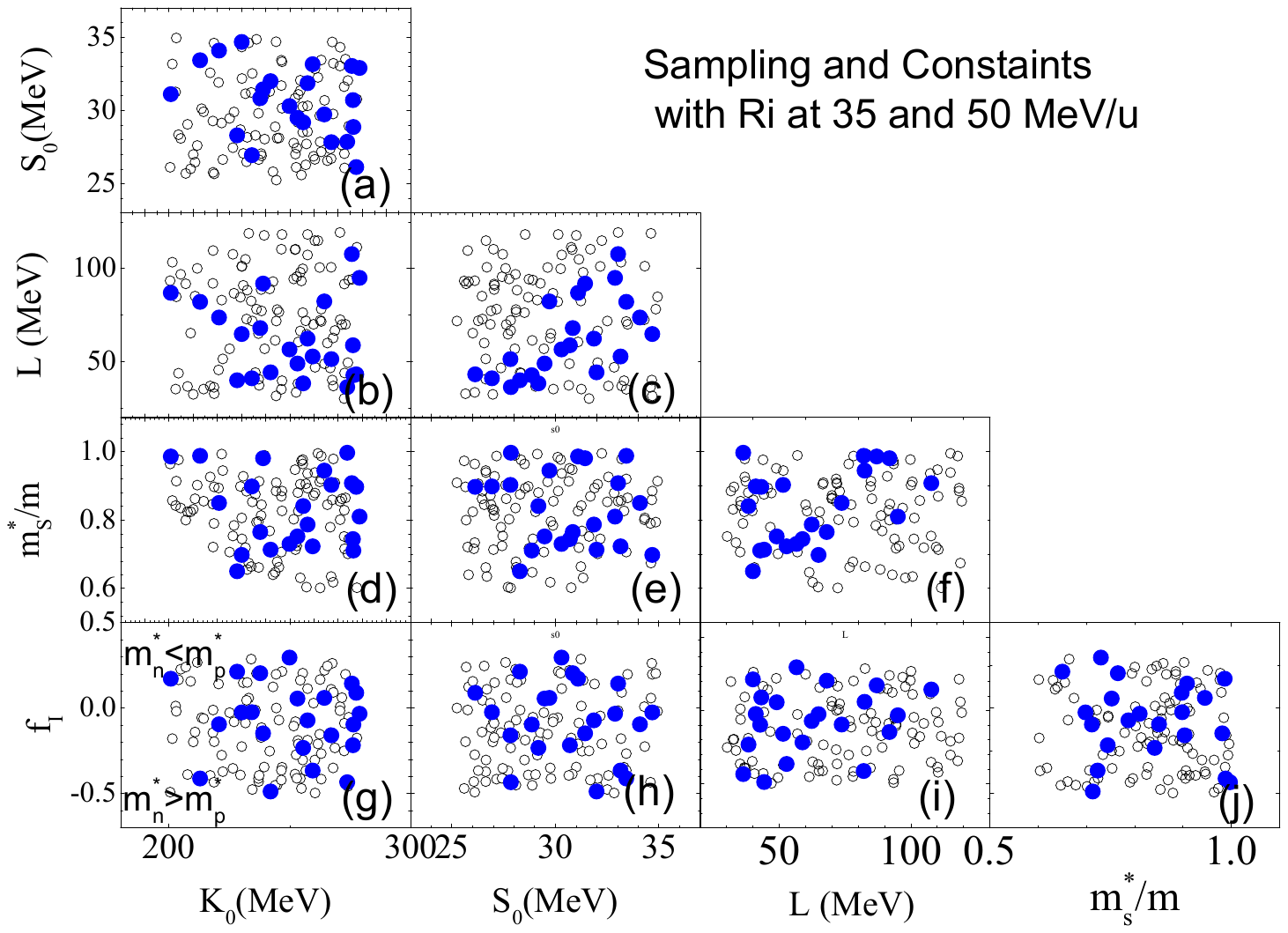}
\setlength{\abovecaptionskip}{0pt}
\caption{\label{samp-120}(Color online) Sampled parameter set points in 5-D parameter space, blue solid points are the sets which can reproduce two isospin diffusion data.}
\setlength{\belowcaptionskip}{0pt}
\end{figure}

\emph{Results and Discussions}.
We choose commonly used values of $\rho_0=0.16fm^{-3}$, $E_0=-16MeV$, and thus only 5 nuclear matter parameters $K_0$, $S_0$, $L$, $m_s^*$, $f_I$, are left in the parameter space. The different parameter sets correspond to the different points in 5 dimension parameter space, $x =\{K_0, S_0, L,m_s^*,f_I\}$. We sampled 120 points in 5D parameter space in the range which we listed in Table~\ref{tab:table1} under the condition that $\eta\ge 1.1$. $\eta\ge 1.1$ is used for guaranteeing the reasonable three-body force in the transport model calculations. The range of these nuclear matter parameters are chosen based on the $prior$ information of Skyrme parameters (Supplementary Fig. 1). As an example, the 120 sampled points are presented as open and solid circles in two-dimensional projection in Figure~\ref{samp-120}. The points of parameter sets uniformly distribute in two-dimensional projection except for the plots of $K_0$ and $m^*_s/m$ due to the restriction of $\eta\ge1.1$. We perform the calculations for isospin transport diffusion at 35 MeV/u and 50 MeV/u at b=5-8fm with the impact parameter smearing\cite{Lili18} for $^{112,124}$Sn+$^{112,124}$Sn. 10,000 events are calculated for each point in the parameter space and simulation are stopped at 400fm/c. The calculations are performed on TianHe-1 (A), the National Supercomputer Center in Tianjin.

\begin{table}[b]%The best place to locate the table environment is directly after its first reference in text
\caption{\label{tab:table1}%
Model parameter space used in the codes for the simulation of $^{112,124}$Sn+$^{112,124}$Sn reaction. 120 parameter sets are sampled in this space by using Latin Hyper-cuber Sampling method.}
\begin{ruledtabular}
\begin{tabular}{lcc}
\textrm{Para. Name}& \textrm{Values}& \textrm{Description}\\
%\multicolumn{1}{c}{\textrm{Decimal}}&

\colrule
$K_0$ (MeV) & [200,280] & Incompressibility \\
$S_0$ (MeV) & [25,35] & Symmetry energy coefficient \\
$L$ (MeV) & [30,120] & Slope of symmetry energy \\
 $m_s^*/m_0$ & [0.6,1.0] & Isoscalar effective mass \\
$f_I=(\frac{m_0}{m_s^*}-\frac{m_0}{m_s^*})$ & [-0.5,0.4] & $f_I=\frac{1}{2\delta}(\frac{m_0}{m_n^*}-\frac{m_0}{m_p^*})$ \\
\end{tabular}
\end{ruledtabular}
\end{table}

\begin{figure}[htbp]
\centering
\includegraphics[angle=0,scale=0.28]{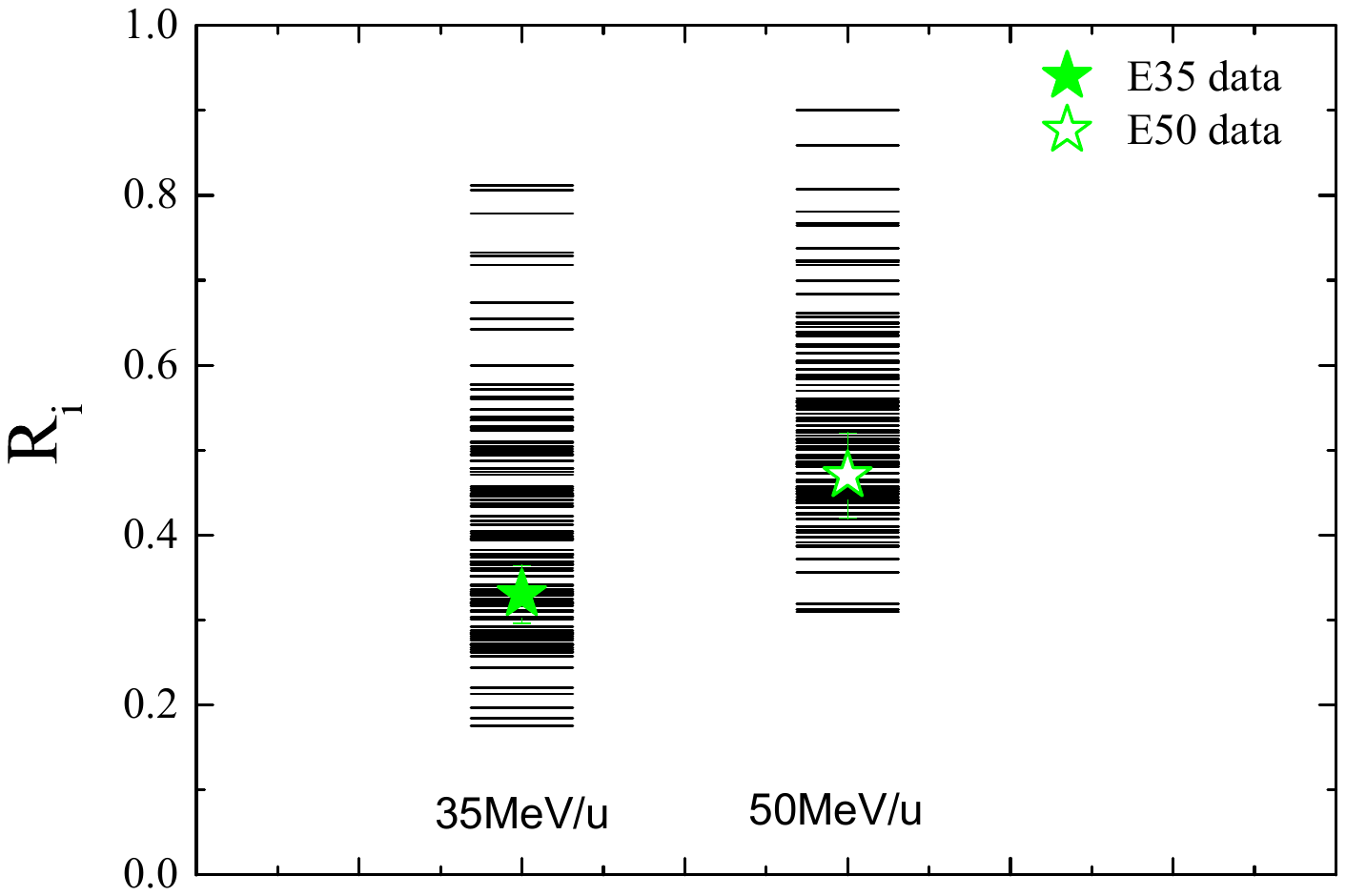}
\setlength{\abovecaptionskip}{0pt}
\caption{\label{ri-theexp}(Color online) Stars are the isospin diffusion data at 35 MeV/u and 50 MeV/u \cite{Sun10, TXLiu07}, lines are the calculated isospin transport ratios with 120 parameter sets.}
\setlength{\belowcaptionskip}{0pt}
\end{figure}

In Figure~\ref{ri-theexp}, the lines represent the calculated results of isospin transport ratio $R_i$ with 120 parameter sets. The isospin transport ratios $R_i$ is defined as
\begin{equation}
R_i=\frac{2X_{ab}-X_{aa}-X_{bb}}{X_{aa}-X_{bb}}
\label{ridef}
\end{equation}
which is constructed from at least three reaction systems, two symmetric systems, such as $^{112}$Sn+$^{112}$Sn and $^{124}$Sn+$^{124}$Sn, and one mixing system $^{112}$Sn+$^{124}$Sn or $^{124}$Sn+$^{112}$Sn. In Eq.(\ref{ridef}), $a=^{124}Sn$, $b=^{112}Sn$ and $X=\delta$ which is the isospin asymmetry of emitting source\cite{Tsang09,Zhang12} in the transport model calculations. In theory, the definition on the emitting source comes from the physical process where the isospin diffusion reflects the isospin asymmetry of the projectile-like residue immediately after the collision and prior to secondary decay. Based on this concept, the `emitting source' are constructed from the emitted nucleons and fragments with velocity greater than half of the beam velocity, i.e. $v_i>0.5v_b^{c.m.}$, $i$=fragments, nucleons. The values of isospin transport ratio at projectile region reflect the isospin diffusion which depends on the stiffness of symmetry energy and the strength of effective mass\cite{LWChen05,Zhang15}.

Two stars in middle panel of Figure~\ref{ri-theexp} are the experimental data which are isospin transport ratio at projectile rapidity region, constructed from the isoscaling parameter $X=\alpha_{iso}$ at 50 MeV/u\cite{Tsang04,TXLiu07} and the ratio of $X=Y(^7Li)/Y(^7Be)$\cite{Sun10, TXLiu07} at the beam energy of 35 MeV/u. The isospin transport ratio obtained from $\delta$ can be compared to that constructed from $\alpha_{iso}$ or $Y(^7Li)/Y(^7Be)$ assuming there is a linear relationship between them\cite{TXLiu07}. As shown in Figure~\ref{ri-theexp}, the full model calculated results show large spread around the experimental data\cite{Tsang04,TXLiu07,Sun10}. By comparing the calculations to the data, we find 22 parameter sets that can reproduce the isospin diffusion data within experimental errors.% which are compiled in the supplementary.

We highlight those points that can reproduce the experimental data within experimental errors with blue solid symbols in upper panel of Figure~\ref{samp-120}. Generally, one can observe: 1) $L$ increase with $S_0$. The constrained points distribute in the bottom-right corner in the $S_0$-$L$ plot (panel (c)), and the large $L$ with small $S_0$ are ruled out. From panel (j), the isospin data is not sensitive to the effective mass. This parameter seems to be more sensitive to double neutron and proton spectral ratios\cite{Pierre19}.

%\begin{table}[htbp]
%\caption{\label{tab:5dnmp} The values of nuclear matter parameter sets that can reproduce isospin diffusion and neutron skin of $^{208}Pb$ ($\Delta R_{np}=0.15\pm0.02$fm). (+) means the sets can produce the data, (-) means not.}
%    \begin{tabular}{ccccccc}
%    \hline
%    \hline
%$K_0$ (MeV) & $S_0$ (MeV) & $L$ (MeV) & $m^*_s/m_0$ & $f_I$ & $R_i$ & $\Delta R_{np}$ \\
%\hline
%234.391 & 26.936 & 41.147 & 0.898 & -0.024 & + & + \\
%277.553 & 26.124 & 43.235 & 0.897 & 0.089 & + & + \\
%259.484 & 33.146 & 52.855 & 0.723 & -0.366 & + & +\\
%257.436 & 31.863 & 62.418 & 0.787 & -0.072 & + & -\\
%249.937 & 30.298 & 56.647 & 0.73 & 0.295 & + & +\\
%267.291 & 27.828 & 51.482 & 0.903 & -0.16 & + & + \\
%276.418 & 28.86 & 42.831 & 0.711 & -0.097 & + & +\\
%200.821 & 31.098 & 87.039 & 0.986 & 0.171 & + & -\\
%228.2 & 28.292 & 40.048 & 0.65 & 0.212 & + & +\\
%253.203 & 29.474 & 49.084 & 0.752 & 0.055 & + & +\\
%242.098 & 31.985 & 44.36 & 0.713 & -0.488 & + & +\\
%239.014 & 31.441 & 91.905 & 0.981 & -0.148 & + & -\\
%230.13 & 34.676 & 64.931 & 0.698 & -0.026 & + & -\\
%220.763 & 34.081 & 73.762 & 0.85 & -0.096 & + & -\\
%237.836 & 30.837 & 68.072 & 0.765 & 0.203 & + & -\\
%276.165 & 30.705 & 58.846 & 0.744 & -0.218 & + & +\\
%212.881 & 33.425 & 82.13 & 0.988 & -0.413 & + & -\\
%273.816 & 27.854 & 36.382 & 0.997 & -0.435 & + & +\\
%278.918 & 32.888 & 95.046 & 0.81 & -0.033 & + & -\\
%255.597 & 29.184 & 38.419 & 0.841 & -0.233 & + & +\\
%275.783 & 33.03 & 107.768 & 0.908 & 0.143 & + & -\\
%264.335 & 29.718 & 82.428 & 0.945 & 0.059 & + & -\\
%    \hline
%    \hline
%    \end{tabular}
%\end{table}

By using the Eq.~(\ref{eq:skyqmd}) and relations in reference \cite{Zhang06,Zhang14}, we can construct the effective standard Skyrme parameter sets, \{$t_0$, $t_1$, $t_2$, $t_3$, $x_0$, $x_1$, $x_2$, $x_3$, $\sigma$\}, except the coefficient related to the spin-orbit terms. In table~(\ref{tab:22skyrmes}), we present the extracted 22 standard Skryme parameter sets and the corresponding neutron skin of $^{208}Pb$ i.e., $\Delta R_{np}\equiv <r^2_n>^{1/2}-<r^2_p>^{1/2}$, based on the RDV method. The averaged values of neutron skin of $^{208}Pb$ is $\delta R_{np}=0.18\pm0.04$ fm, and it is consistent with the neutron skin values extracted from the experiments from reference \cite{Hoff80, Trzc01, Klos07, Klim07, Zeni10, Tamii11, Abra12, Piek12, Tarb14, Roca15,Tsang12}.
%where $\delta R_{np}=0.18\pm0.027$ fm. By comparing with the current published neutron skin data of $^{208}$Pb\cite{Hoff80, Trzc01, Klos07, Klim07, Zeni10, Tamii11, Abra12, Piek12, Tarb14, Roca15,Tsang12}, we find all the 22 sets predict the neutron skin values within the data errors. If we assume the neutron skin values as $\Delta R_{np}=0.15\pm0.02$fm, 12 of them can produce both isospin diffusion and neutron skin and 8(4) of them are $f_I<0$, i.e. $m^*_n>m^*_p$ ($f_I>0$, or $m^*_n<m^*_p$). All of the parameter sets that can reproduce isospin diffusion and neutron skin of $^{208}Pb$ in the assume range ($\Delta R_{np}=0.15\pm0.02$fm) are listed in Table.\ref{tab:5dnmp}.

\begin{table*}[htbp]
\caption{\label{tab:22skyrmes} Extracted 22 standard Skyrme parameter sets and the corresponding neutron skin values based on RDV method. $t_0$ in $MeVfm^3$, $t_1$ and $t_2$ in $MeVfm^5$, $t_3$ in $MeVfm^{3\sigma+3}$, $x_0$ to $x_3$ is dimensionless quantities. In the RDV calculations of this work, $W_0=130 MeVfm^{5}$ and $\rho_0=0.16fm^{-3}$.}
    \begin{tabular}{cccccccccc}
    \hline
    \hline
$t_0$ & $t_1$  & $t_2$  & $t_3$ & $x_0$ & $x_1$ & $x_2$ & $x_3$ & $\sigma$ &$\Delta R_{np}$ \\
\hline
-1890.80 &427.97 &-490.81 &	12571.72 &	0.10669 &	-0.19396 &	-0.7161 &	0.15416 &	0.29804 &	0.14356\\
-1374.17 &428.19 &-607.42 &	10814.29 &	0.04292 &	-0.26258 &	-0.81939 &	0.24329 &	0.51892 &	0.14216\\
-1569.42&	474.60&	3.93&	9415.46&	0.21035&	-0.03708&	-41.13867&	-0.02844&	0.37265&	0.16947\\
-1572.00&	455.14&	-359.50&	10186.44&	0.10568&	-0.18487&	-0.69112&	0.07323&	0.38608&	0.1803\\
-1714.97&	472.30&	-688.83&	10110.07&	0.34791&	-0.39789&	-1.01437&	0.97341&	0.31666&	0.15358\\
-1452.20&	426.91&	-352.89&	10979.89&	-0.02416&	-0.11056&	-0.50064&	-0.25793&	0.46733&	0.16337\\
-1395.03&	478.63&	-263.07&	8737.27&	0.20269&	-0.18678&	-0.68719&	0.48667&	0.47509&	0.13906\\
-3048.33&	410.78&	-744.73&	19381.38&	-0.28089&	-0.3043&	-0.8462&	-0.35056&	0.16036&	0.22825\\
-3312.92&	501.46&	-515.21&	17988.52&	1.00059&	-0.36089&	-1.06232&	1.48966&	0.10376&	0.10553\\
-1644.99&	465.37&	-460.59&	10070.75&	0.24038&	-0.26259&	-0.86375&	0.55912&	0.34745&	0.14974\\
-1914.52&	477.95&	140.60&	10865.66&	0.15117&	0.02588&	-2.31398&	-0.12133&	0.25704&	0.15966\\
-1766.26&	411.68&	-411.04&	12629.01&	-0.43493&	-0.10372&	-0.52328&	-0.93988&	0.34248&	0.23979\\
-2480.04&	483.17&	-323.15&	13757.39&	0.39189&	-0.22784&	-0.82337&	0.54526&	0.16807&	0.17542\\
-2359.49&	438.85&	-383.47&	14591.08&	-0.02704&	-0.15899&	-0.63047&	-0.17633&	0.20869&	0.20927\\
-1945.23&	461.46&	-625.89&	11613.88&	0.15946&	-0.34378&	-0.95171&	0.41995&	0.26249&	0.17595\\
-1393.55&	467.84&	-169.89&	9181.01&	0.06398&	-0.11247&	-0.2356&	-0.13504&	0.48318&	0.16756\\
-2406.93&	410.43&	-139.80&	15831.81&	-0.50854&	0.06498&	0.90667&	-1.02398&	0.21879&	0.24224\\
-1396.68&	408.85&	-121.72&	11646.34&	0.0986&	0.08113&	1.25635&	-0.43832&	0.51157&	0.15322\\
-1368.43&	448.90&	-418.69&	9938.92&	-0.21753&	-0.20303&	-0.73659&	-0.6341&	0.51343&	0.22588\\
-1579.20&	441.03&	-234.77&	10767.52&	0.19335&	-0.08009&	-0.25897&	0.09028&	0.39256&	0.15074\\
-1386.09&	425.85&	-670.47&	10922.75&	-0.33682&	-0.29417&	-0.85204&	-0.68126&	0.51127&	0.2551\\
-1474.21&	418.40&	-605.68&	11375.98&	-0.25086&	-0.23842&	-0.78177&	-0.4952&	0.45917&	0.21354\\
    \hline
    \hline
    \end{tabular}
\end{table*}

%\begin{figure}[htbp]
%\centering
%\includegraphics[angle=0,scale=0.28]{fig2b-nskin-pb208-1017.pdf}
%\includegraphics[angle=270,scale=0.3]{fig1.eps}
%\setlength{\abovecaptionskip}{0pt}
%\caption{\label{nskin}(Color online) Symbols are the data of $\Delta R_{np}$ obtained from different groups, lines are the results of $\Delta R_{np}$ obtained with 22 parameter sets with the restricted density variational method.}
%\setlength{\belowcaptionskip}{0pt}
%\end{figure}

Based on the extracted 22 Skyrme parameter sets, we can also obtain the corresponding symmetry energy which is a hot topic in the physics of heavy ion collisions. The form of corresponding density dependence of symmetry energy for cold nuclear matter in the Skyrme-Hartree-Fock approach read as,
\begin{eqnarray}
S(\rho)&=&\frac{1}{3}\frac{\hbar^2}{2m}(\frac{3\pi^2}{2})^{2/3}\rho^{2/3}-\frac{1}{8}t_0(2x_0+1)\rho\\\nonumber
&&-\frac{1}{24}(\frac{3\pi^2}{2})^2(3\Theta_v-2\Theta_s)\rho^{5/3}\\\nonumber
&&-\frac{1}{48}t_3(2x_3+1)\rho^{\sigma+1}\\\nonumber
&=&\frac{1}{3}\frac{\hbar^2}{2m}(\frac{3\pi^2}{2}\rho)^{2/3}+(A_{sym}u+B_{sym}u^\eta+C_{sym}u^{5/3}),
\end{eqnarray}
where $u=\rho/\rho_0$.  $S_0=S(\rho_0)$ and
\begin{eqnarray}
L&=&3\rho_0\frac{\partial S(\rho)}{\partial \rho}|_{\rho_0}\\\nonumber
&=&\frac{2}{3}\frac{\hbar^2}{2m}(\frac{3\pi^2}{2})^{2/3}\rho_0^{2/3}-\frac{3}{8}t_0(2x_0+1)\rho_0\\\nonumber
&&-\frac{5}{24}(\frac{3\pi^2}{2})^2(3\Theta_v-2\Theta_s)\rho_0^{5/3}\\\nonumber
&&-\frac{3(\sigma+1)}{48}t_3(2x_3+1)\rho_0^{\sigma+1}.
\end{eqnarray}

\begin{figure}[htbp]
\centering
\includegraphics[angle=0,scale=0.35]{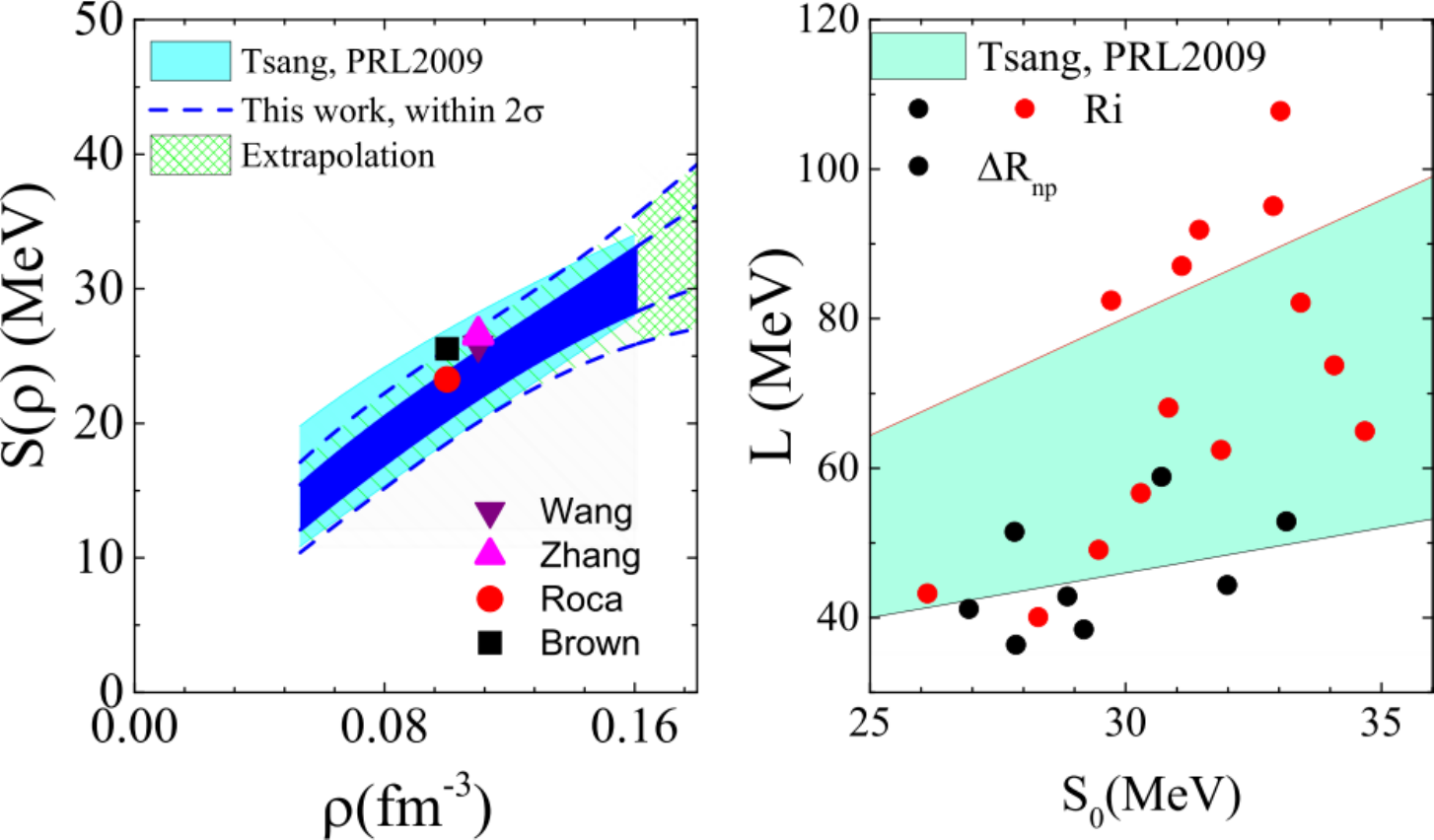} %{fig3-srho-ri3550new.pdf}
\setlength{\abovecaptionskip}{20pt}
\caption{\label{srho-ri}(Color online). Symmetry energy information from 22 Skyrme parameter sets. Left panel is for density dependent of symmetry energy in the range of 1/3-1.2 normal denisty. Right panel is for the $S_0$ and $L$ values. Black points in middle and right panel are the sets can reproduce the assumed neutron skin thickness of $^{208}Pb$, i.e. $\delta R_{np}=0.15\pm0.02$ fm.}
\setlength{\belowcaptionskip}{20pt}
\end{figure}

The density dependence of the symmetry energy obtained from 22 parameter sets are presented in left panel of Figure~\ref{srho-ri}. The shadow region with blue color represents for the $S(\rho)$ constrained from the two isospin diffusion data, i.e., $R_i$ at 35 MeV/u and 50 MeV/u, within 1$\sigma$. The region within the blue dashed lines are the constrained $S(\rho)$ within 2$\sigma$ uncertainties. The shadow region with cyan color is the constraint of symmetry energy obtained in 2009 by analyzing the data of isospin diffusion, isospin transport ratio, and double neutron to proton yield ratio at 50 MeV/u with ImQMD codes\cite{Tsang09}, where the corresponding density dependence of symmetry energy is
\begin{equation}
S(\rho)=\frac{1}{3}\frac{\hbar^2}{2m}(\frac{3\pi^2}{2})^{2/3}\rho^{2/3}+\frac{C_s}{2}(\frac{\rho}{\rho_0})^\gamma.
\end{equation}
Compare to the constraints of $S(\rho)$ by 2009 HIC data, the new analysis improve the constraints at the density below $\sim 0.13 fm^{-3}$ because we include isospin diffusion data at 35 MeV/u in this analysis. The uncertainties of the constraints of symmetry energy around normal density become larger than that in 2009, because the current analysis includes the uncertainties of $K_0$, $m^*_s$, and $f_I$. The symmetry energy obtained from the electric dipole polarizability in $^{208}$Pb \cite{ZZhang14}(red circle), properties of double magic nuclei and masses of neutron-rich nuclei\cite{Brown13} (black square and up triangle) and Fermi-energy difference in finite nuclei\cite{Wang13} (blue down triangle) are also presented in the left panel of Figure~\ref{srho-ri}. The symmetry energy obtained in this work is consistent with the previous constraints within 2$\sigma$ uncertainties.

The consistence of the symmetry energy obtained from 22 Skyrme parameter sets and the symmetry energy constraints from other nuclear structure studies\cite{ZZhang14,Brown13,Wang13} is because both contain the information of symmetry energy at subsaturation density. It can be simply understood from the right panel of Figure~\ref{srho-ri} by using the approach of sensitive density proposed by W.G. Lynch and M.B. Tsang\cite{Lynch18}. %For simplicity, we only present the constrained values of $S_0$ and $L$ on the $S_0$ and $L$ plane by comparing to HIC data with circles.
The shaded region in the right panel of Figure~\ref{srho-ri} is the constraint given by reference \cite{Tsang09}, and the points in the right panel are the constraint by the isospin diffusion data at 35 and 50 MeV/u in this work. The correlation between $S_0$ and $L$ is consistent with our previous work\cite{Pierre19}.
By best linear fitting these points, the values of $\frac{\partial S_0}{\partial L}$ can be obtained, and we got $\frac{\partial S_0}{\partial L} = 0.061$ with standard error 0.022. Thus, the corresponding sensitive density is $\rho_s/\rho_0=0.685-0.946$ with $2\sigma$ of the $\frac{\partial S_0}{\partial L}$. The range of sensitive density is consistent with the dynamical prescription of isospin diffusion process in peripheral heavy ion collisions, where the density in the neck region evolves from normal density to subnormal density until the neck breaks. This is also close to the corresponding average density region in the nuclear skin studies\cite{Khan12,Brown13}.

We do not use the data of double neutron to proton yield ratios\cite{Famiano06} to extract the effective Skyrme energy density functional in this work, because the data of double neutron to proton yield ratios in 2006\cite{Famiano06} has large errors and later proved to be wrong. In addition, the analysis from both QMD type or BUU type models could not well reproduce the neutron and proton yields due to the inadequacy of mechanism in describing the light particle formation, especially at the beam energy of 50 MeV/u. Analysis of the single and double coalescence invariant neutron to proton yield ratios at 120 MeV/u can be found in \cite{Pierre19}.

\emph{Summary}. In summary, we established 22 Skyrme parameter sets by comparing, the isospin diffusion data at 35 and 50 MeV/u, to transport model calculations where we use the nuclear matter parameter as an input for removing the $prior$ correlation between them. Based on the restricted density variation method, the neutron skin of $^{208}Pb$ from 22 parameter sets is $\Delta R_{np}=0.18\pm0.04$fm. The values of {$K_0$, $S_0$, $L$, $m_s^*/m$, and $f_I$} of 22 Skyrme parameter sets are estimated. Our calculations show the positive correlation between $S_0$ and $L$ under the constraints from isospin diffusion data, and the $L$ values obtained from 22 parameter sets distribute from 30 to about 100 MeV. Most of the estimated values of $f_I$ in this work are negative which corresponds to the $m^*_n>m^*_p$, but we can not rule out $f_I>0$ (i.e. $m^*_n<m^*_p$) by using the diffusion data. Future works on comprehensive test of Skyrme parameter sets with nuclear structure, neutron stars as well as in heavy ion collision will be helpful for constraining the isospin asymmetric equation of state over a large density region.
%Finally, the standard Skyrme potential energy density functional parameters are obtained by the expressions given in Eq.~\ref{eq:skyqmd} with the extracted values of the values of \{$K_0$, $S_0$, $L$, $m_s^*/m$, $f_I$\} ,which are provided in supplementary.

\begin{acknowledgements}
The authors thanks for the helpful discussions with Prof. M. B. Tsang and Prof. H. St\"{o}cker, and C. Y. Tsang. This work was supported by the National Science Foundation of China Nos.11875323, 11875125, 11475262, 11790323, 11790324, and 11790325, the National Key R\&D Program of China under Grant No. 2018 YFA0404404 and the Continuous Basic Scientific Research Project (No. WDJC-2019-13). The work was carried out at National Supercomputer Center in Tianjin, and the calculations were performed on TianHe-1 (A).
\end{acknowledgements}

\end{document}